# Synthesis and characterization of vertically aligned La$_{0.7}$Sr$_{0.3}$MnO$_3$:NiO nanocomposite thin films for spintronic applications

Gyanendra Panchal[1,2,*], Anjali Panchwanee[2], Manish Kumar[3], Katharina Fritsch[1], Ram Janay Choudhary[2,*], and Deodutta Moreshwar Phase[2]

[1]*Department Methods for Characterization of Transport Phenomena in Energy Materials, Helmholtz-Zentrum Berlin für Materialien und Energie, D-14109 Berlin, Germany*
[2]*UGC DAE Consortium for Scientific Research, University Campus, Khandwa Road, Indore 452001, India*
[3]*Pohang Accelerator Laboratory, POSTECH, Pohang 37673, South Korea*

The microstructures and interfaces of two-phase vertically aligned nanocomposite (VAN) thin films play a key role in the design of spintronic device architectures and their multifunctional properties. Here, we show how the microstructures in self-assembled VAN thin films of La$_{0.7}$Sr$_{0.3}$MnO$_3$:NiO (LSMO:NiO) can be effectively tuned from nano-granular to nano-columnar, and to nano-maze by controlling the number of laser shots from the two constituent phase targets in the pulsed laser deposition (PLD) film growth. The observed microstructural induced strain is found to significantly enhance the magnetoresistance in a very broad temperature range between 10-240 K and to modulate the in-plane exchange bias (EB), with the largest EB value observed in the maximally strained heterostructures. Most interestingly, a unique perpendicular exchange bias (PEB) effect is also observed for these heterostructures with an enhanced PEB field of up to 230 Oe. X-ray magnetic circular dichroism and training effect measurements demonstrate that the observed EB is disorder-induced and arises due to the pinning of NiO uncompensated moments at the disordered interface which is ferromagnetically coupled with LSMO. Furthermore, systematic changes in the electronic structure across the vertical interface related to a variation of the Mn$^{3+}$/Mn$^{4+}$ content arise as a consequence of out-of-plane tensile strain.

**E-mail:** gyanendra.panchal@helmholtz-berlin.de; ramjanay@gmail.com

## Introduction:

Strain engineering has been extensively used as an effective approach to manipulate the multifunctional properties of thin films.[1,2] Such heterostructures in conventional *horizontal* interfacial layered geometries in bilayer, multilayer or superlattice arrangements composed of magnetically active transition metal oxides exhibit a plethora of interesting phenomena such as interfacial ferromagnetism,[3-5] anomalous Hall-effect,[6,7] or interfacial superconductivity.[8] In such epitaxial planar heterostructures, the lateral in-plane strain derived from lattice mismatch is, however, limited up to a certain critical thickness because of strain relaxation through misfit dislocations, oxygen vacancies or other defects.[9]

In recent years, epitaxial *vertically* aligned nanocomposite (VAN) thin films have been found to provide a new playground to enhance the physical properties and to yield new functionalities via interplay among charge, orbital and spin degrees of freedom.[10–15] In these VAN thin films, functional properties can be systematically manipulated via the choice of materials, vertical strain, lattice mismatch between the two-phase system and the substrate, and more importantly by tailoring the size, shape, orientation, and phase fraction of the constituent phases.[16] Therefore, such VAN thin films prove advantageous over conventional layered structures because the modulation in the microstructure of composite thin films provides a wide range for tuning functional properties ranging from electronic structure, magneto-electric properties,[17–19] exchange bias,[20,21] electrical transport,[22–24] spontaneous polarization,[25] and magnetic anisotropy.[26] Among these, exchange bias and low-field magnetoresistance are crucial parameters for next generation magnetic storage devices such as read heads with high thermal stability and reduced dimension and their potential application in future spintronics applications.[27]

The bilayer and the superlattices approaches have been earlier studied by introducing room temperature ferromagnetic (FM) La$_{0.7}$Sr$_{0.3}$MnO$_3$(LSMO) thin films along with a layer of different functional oxide materials such as LaNiO$_3$, BiFeO$_3$, BaTiO$_3$, NiO, ZnO, SrMnO$_3$ and MgO.[28–33] In these planar heterostructures, the possibility of interfacial tuning is rather limited due to a fixed interfacial contact area between the two materials. In the case of LSMO based vertical heterostructures, the different competing exchange interactions (mediated by Mn ions), charge transfer and oxygen vacancies/defects at the interfaces can be tuned via vertical microstructure, which plays a key role in the control and manipulation of their physical properties, especially their magnetoresistance and exchange bias effect.[34–36] Recently, Jijie Huang et al.[37] showed the unique perpendicular exchange bias properties in LSMO:NiO VAN nanocomposite thin film grown on a flexible mica substrate.

In this work, we present a comprehensive study of the two-phase self-assembled nanocomposite LSMO:NiO hybrid system, of which we have grown thin films with different microstructure on single crystalline LaAlO$_3$ (001) and SrTiO$_3$ (001) substrates by pulsed laser deposition (PLD). NiO has been used as secondary phase to form the self-assembled VAN because of its G-type antiferromagnetic nature with fairly high Néel tempe-rature of $T_N$ =525 K and because of its large lattice mismatch with the ferromagnetic LSMO matrix (a$_{NiO}$ = 4.18 Å, a$_{LSMO}$ (001) =3.87 Å). We have studied the structural, magnetic, magnetotransport and electronic structure properties of these self-assembled VAN nanocomposite thin films. Our results show that fine-tuning of the microstructure-induced out-of-plane (OP) strain and interfacial disorder/ and grain boundaries plays a crucial role to effectively modify the bulk and microscopic magnetic properties, electrical transport and the electronic structural properties of these LSMO:NiO VAN thin films.

**Experimental:**

*Nanocomposite Films Growth:*

To prepare nanocomposite thin films of LSMO and NiO, two separate single-phase targets were alternately used. A NiO target is prepared by using commercial Ni(II)-oxide powder (by Sigma-Aldrich) and a LSMO target was prepared by using conventional solid state route in stoichiometric ratios from La$_2$O$_3$, Sr(CO$_3$)$_2$, and MnO$_2$ to yield La$_{0.7}$Sr$_{0.3}$MnO$_3$. Nanocomposite thin films of LSMO:NiO were simultaneously deposited on (001)-oriented single crystalline LaAlO$_3$(LAO) and SrTiO$_3$(STO) substrates, by PLD using a KrF excimer laser with a wavelength of 248 nm. In particular, two semi-circular targets of NiO and LSMO were used for deposition as shown in Figure 1. During the deposition, the substrate temperature was kept at 750 $^o$C, the oxygen partial pressure (OPP) in the chamber was 250 mTorr and the energy flux at the target was kept at 2.0 J/cm$^2$ with a 3 Hz repetition rate for both materials. Prior deposition, the substrates were ultrasonically cleaned with acetone followed by methanol to remove any contamination or oil on the substrate surface. After the deposition process, the films were cooled in 375 Torr oxygen pressure with a cooling rate of 5 $^o$C/min.

*Characterizations:*

The structural properties of these VANs were determined by X-ray diffraction measurements using a Bruker D2 PHASER setup with Cu K-$\alpha$ lab source, collecting data in the $\theta$-$2\theta$ geometry. The microstructure of the films was studied using a field emission scanning electron microscope (FE-SEM NOVA Nano SEM-450 by FEI). DC magnetization measurements were performed in a 7-Tesla SQUID-vibrating sample magnetometer (SVSM by Quantum design, Inc., USA) from 5 to 355 K. Room temperature X- ray absorption near-edge structure (XANES) measurements at La $M$ as well as Ni, Mn $L$-edges and O $K$-edges were performed in total electron yield (TEY) mode under $10^{-10}$ Torr vacuum at polarized light soft X-ray absorption spectroscopy beamline BL-01, Indus-2, RRCAT, India. Temperature-dependent (80 - 320 K) X-ray magnetic circular dichroism (XMCD) measurements under 6 kOe magnetic field were performed at the Pohang Accelerator Laboratory (PAL) Korea on the 2A MS undulator beamline. Magneto-electric transport properties in the temperature range 5-300 K were measured by four-probe method using custom-built setup along with an 80 kOe superconducting magnet from Oxford instruments, UK.

## Results and discussion:

### Structural properties

The deposition process of the composite thin films, is encoded by the formula [(LSMO)s/(NiO)s]t where 's' is the number of laser shots and 't' is the deposition time in minutes. The total deposition time 't' is kept at 50 minutes for all composite thin films studied and only the number of laser shots 's' is used to tune the microstructure of the composite thin films. The complete schematic for the synthesis of the thin films is shown in Figure 1. By controlling the rotation speed of the target to adjust the number of laser shots, we have simultaneously grown three sets of composite thin films on LAO/STO substrates with different microstructures labelled as LA7/ST7, LA14/ST14 and LA25/ST25 where s =7, s =14 and s = 25, respectively.

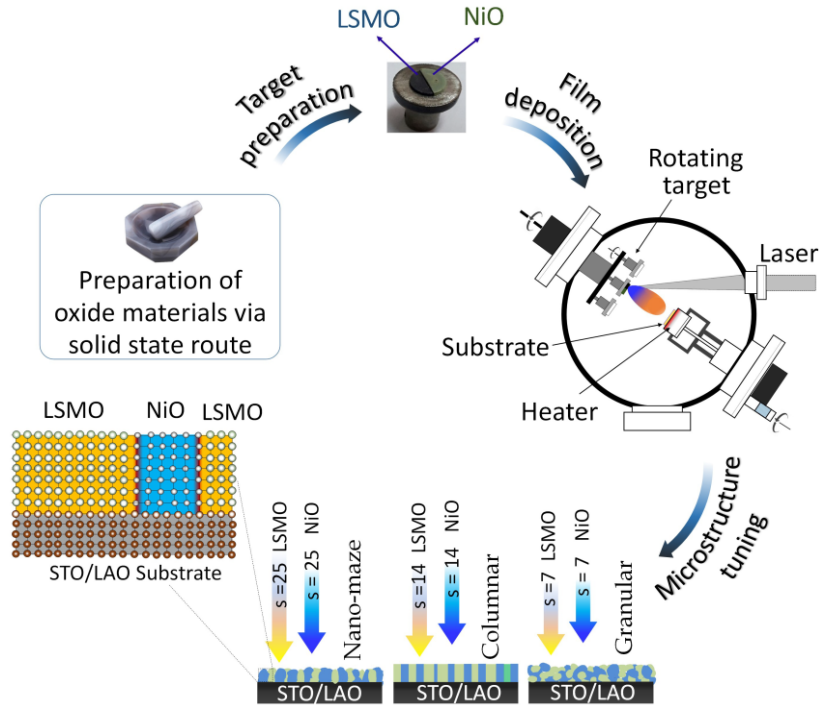

Figure 1: Scheme for the fabrication of LSMO:NiO nano-composite thin films with different resulting microstructures based nano-granular, nano columnar and nano-maze nanostructures, tuned by the growth parameter: number of laser shots 's' [(LSMO)s/(NiO)s]t(50 minutes).

The volume fractions of LSMO and NiO in the composite thin films are equal due to the same deposition time, but a lower ablation yield of NiO results in the effective thicknesses for NiO and LSMO of 50 ± 5 and 115 ± 5 nm, respectively. X-ray diffractograms of self-assembled nanocomposite thin films LA7/ST7, LA14/ST14 and LA25/ST25 along with those of the bare LAO (001)/STO (001) substrates are shown in Figure 2. For all samples, only reflections attributed to NiO (200) and LSMO appear in addition to the substrate reflections and we conclude that our films are devoid of any impurity phases. Thin films prepared on LAO substrates exhibit a preferred (00$l$) orientation of LSMO and a prominent (002) reflection of NiO.

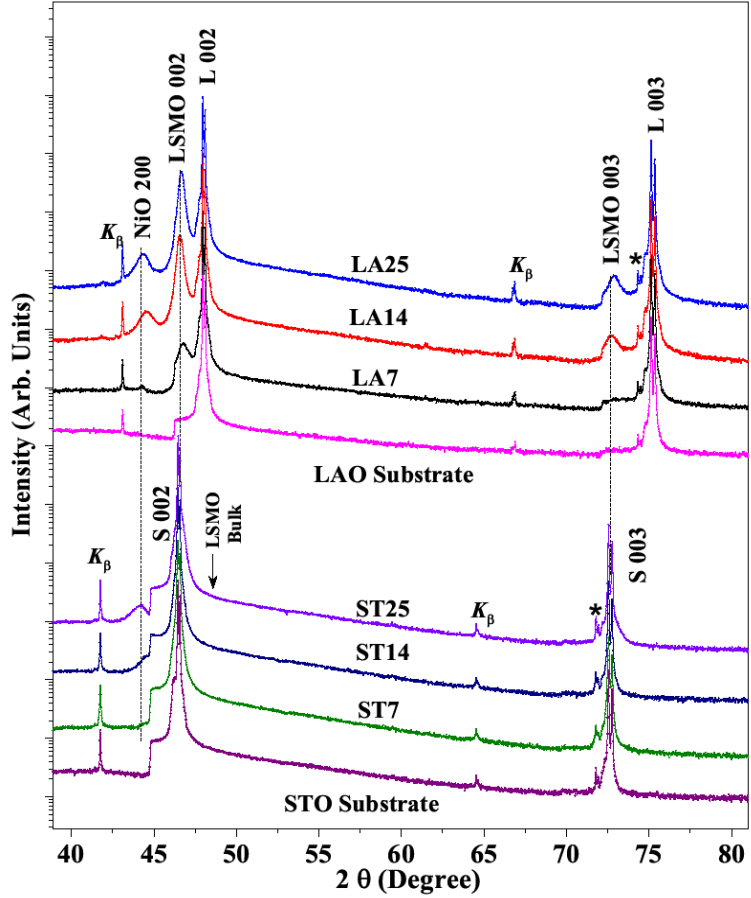

Figure 2: θ-2θ X-ray diffractograms of LSMO:NiO composite thin films LA7/ST7, LA14/ST14 and LA25/ST25 along with those of the reference (001) oriented single crystalline LAO(L)/STO(S) substrates. The peak represented by the asterisk arises due to the Kα satellite interference.[38]

In the case of thin films grown on STO substrates, the (00$l$) reflections of LSMO are smeared out under the (00$l$) reflections of the single crystalline STO due to large vertical strain tuning of LSMO lattice parameters through the strain imparted by the immiscible NiO secondary phase in the composite thin film. Generally, the (002) reflection of LSMO appears on the right hand side of the (002) STO reflection.[29] For the reference, the expected bulk LSMO (002) reflections is indicated by the arrow in Figure 2. All these observations indicate a highly c-axis oriented growth of composite thin films as shown in Figure 2. Calculated out-of-plane (OP) lattice parameters and vertical strain [($a_{film}$-$a_{bulk}$) ×100/$a_{bulk}$] in all composite thin films corresponding to the LSMO and NiO phases are tabulated in Table 1.

It is desirable that LSMO (pseudocubic lattice parameter = 3.88Å) grows epitaxially and forms the host matrix for the secondary phase NiO (epiphyte phase), due to a better crystallographic lattice matching with LAO and STO substrates. The epitaxy of thin films of LSMO on LAO/STO for s =14 is published in a recent report[39] which reveals that only the LSMO fraction in the composite films grows report[39] which reveals that only the LSMO fraction in the composite films grows epitaxially on LAO/STO, whereas the growth of the NiO phase is highly textured. In such composite thin films, the details of formation of horizontal or vertical microstructures additionally depend on several parameters such as surface energy of evaporants and the substrate surface and relative phase fractions of the constituent phases.[40] On the behalf of these consideration, the second phase in this case NiO may not seed its growth on the substrate and tries to align with the host matrix phase (LSMO).[41]

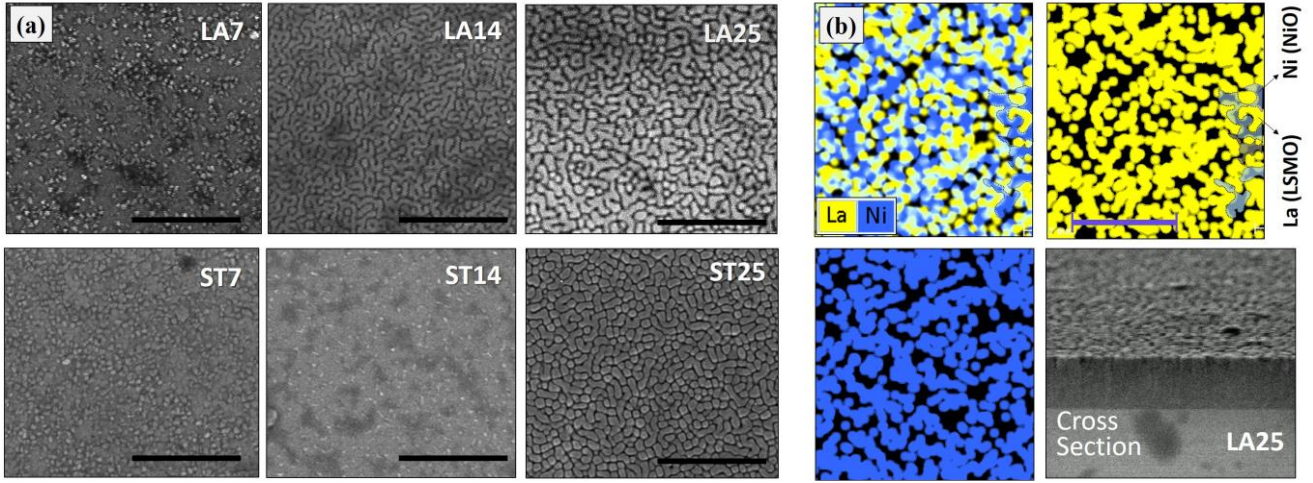

Figure 3: (a) FE-SEM micrographs of LA7/ST7, LA14/ST14 and LA25/ST25 composite thin films grown on single crystalline LAO/STO (001) substrates. (b) EDS elemental mapping of the LA25 sample, distribution of La (yellow) and Ni (blue) elements corresponding to the LSMO and NiO phases along with a cross-sectional view. Horizontal scale-bar in the image is equivalent to 500 nm.

The microstructures of our composite thin films were studied by FE-SEM, revealing different types of microstructures as shown in Figure 3. Usually, the microstructure of composite thin films is controlled by the volume fraction of the secondary phase. Here, we have tuned the microstructure from nano-granular to nano-maze by varying the number of laser shots from s =7 to 25, [Figure 3(a)]. This varying leads to a change in the initial nucleation growth regions.[40] As discussed above, LSMO grows epitaxially and NiO as the epiphyte phase due to their matching and mismatching lattice parameters with respect to substrate, respectively. Because of a smaller fraction of NiO phase reaching the substrate and primary matrix of LSMO, the material initially forms as nano-grains. With an increase in number of laser shots, the volume fraction of NiO increases as well as the strain due to LSMO and substrate, such that the nanocomposite film grows in nano-columnar mode. With further enhancement in volume fraction of NiO, the percolation of the NiO phase take place in the matrix of LSMO causing a maze-like structure.

Therefore, we are able to observe different microstructures for different sets of depositions with the number of laser shots varying between 7, 14 and 25. For s = 7 shots from each target, a nano-granular like microstructure is obtained in the films grown on LAO (LA7) and STO (ST7). When the parameter *s* is increased to *s* = 25, a nano-maze like[42] microstructure is formed in the composite film for both the substrates, LAO (LA25) and STO (ST25) but with larger grain size in LA25 than ST25. In the case of s = 14 shots from each target, a nano-maze and nano-columnar like microstructure are formed on LAO (LA14) and STO (ST14) substrates, respectively. More interestingly, it is observed that larger vertical strain tuning leads to the finest VAN microstructure in the LA14 sample with very fine nano-maze like features and with ultra-fine NiO nano-pillars embedded in the LSMO matrix in the ST14 sample. To study the VAN microstructure, cross-sectional SEM and X-ray energy dispersive spectroscopy (EDS) elemental mappings were carried out on the LA25 sample. Cross-sectional SEM images of other samples are shown in Figure S1 of the Supporting Information. Figure 3(b) shows the SEM micrographs with EDS elemental maps of LA25, the alienated distribution of La (yellow) and Ni (blue) rep- resents the distinct phase separation of LSMO and NiO. This phase separation confirms the vertical heterostructure [shown by arrow in the right side of Figure 3(b)] and indicates the minimal intermixing between matrix and secondary phase with a high quality uniform nano-maze like VAN.

## Magnetic properties

In the VAN system of LSMO:NiO, the magnetization behaviour of LSMO is influenced by (i) strain-induced lattice distortions that elongate $MnO_6$ octahedra and modify the double exchange interaction via modified Mn-O bond length and a modified Mn-O-Mn bond angle,[24] (ii) interfacial magnetic coupling between LSMO and NiO, and (iii) grain boundary defects. Therefore, a VAN microstructure provides a wide tunable range of magnetic properties. The magnetization behaviour as a function of the temperature M(T) of our LSMO:NiO thin films (LA7, LA14 and LA25) of different microstructures in zero field-cooled cooling (ZFC) and field-cooled cooling (FCC) cycles in the temperature range of 5 K- 355 K under an in-plane (IP) applied magnetic field of 50 Oe is shown in Figure 4(a). Sample surface and magnetic field (B) direction during the IP magnetization measurements are shown by a schematic in the bottom inset of Figure 4(a).

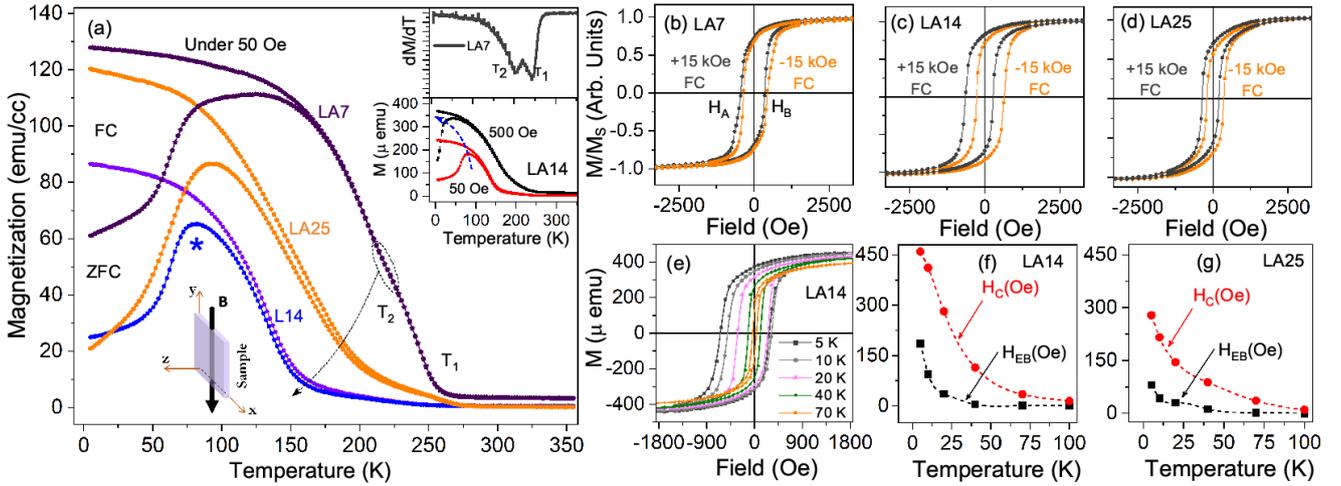

Figure 4: (a) Magnetization versus temperature (M-T) behavior of LA7, LA25 and LA14 composite thin films in zero field-cooled cooling (ZFC) and 50 Oe field-cooled cooling (FCC) protocol. Top inset show the dM/dT of LA7 sample and M-T in FC and ZFC cycles of sample LA14 under applied fields of 50 and 500 Oe. Schematic in bottom inset shows the sample surface and magnetic field (B) direction during the magnetization measurements. (b), (c) and (d) IP magnetic hysteresis loop of sample LA7, LA14 and LA25 recorded at 5 K after +15 kOe and -15 kOe field cooled cooling from 360 K. (e) 1.5 FC M-H of LA14 recorded at temperature 5, 10, 20, 40, 70 and 100 K. Behavior of exchange bias field ($H_{EB}$) and $H_C$ with temperature for LA14 (f) and LA25 (g) sample.

The temperature dependence of the magnetization behaviour of all three composite thin films reveals paramagnetic behaviour at room temperature and ferromagnetic ordering at lower temperatures T < 250 K as well as FC-ZFC divergences. A cusp in M(T) appears in the ZFC cycle [blue star in Figure 4(a)], which moves towards low temperatures when the applied external magnetic field is increased and bifurcates [inset of Figure 4(a)] below the irreversibility temperature. This observation indicates of disordered spin-glass-like behavior, which is discussed in detail for the ST14 sample elsewhere and is not the focus of this work.[39]

The magnetization behavior of the granular composite thin film LA7, is similar to the strained LSMO film with a reduced $T_C$ ~ 243 K ($T_1$) and with a feeble anomaly at $T_2$ around 210 K [(shown by the dashed circle and also visible in the temperature derivative of the magnetization dM/dT shown in the inset of Figure 4(a)]. This granular thin film has the smallest out-of-plane strain of 0.41 %. When the microstructure is tuned from granular into nano-maze (LA14), the OP strain increases to 0.72 %, the Curie temperature is further modified and the anomaly temperature $T_2$ reduces to 135 K. The anomaly temperature $T_2$ changes with microstructure and disappears upon the application of a higher magnetic field of 500 Oe as shown in the inset of Figure 4(a) for the sample LA14. This indicates that it strongly dependents on the competing magnetic interactions and interfacial coupling strength (the higher the strain is, the lower is the value of $T_2$).

Therefore, we deduce that the anomaly at $T_2$ arises due to the modified exchange interaction at the interface. Such kind of anomaly, however at a lower temperature of T=90 K, has also been reported by Ning et al.[43] for $La_{0.7}Ca_{0.3}MnO_3$(LCMO)/NiO thin film heterostructures, and have been attributed to $Mn^{3+}$-O-$Ni^{3+}$ superexchange interactions through a charge transfer effect $Mn^{4+}- Ni^{2+} \rightarrow Mn^{3+}- Ni^{3+}$. Temperatures $T_2$ and $T_1$ are tabulated in Table 1.

TABLE 1: Summary of the parameters describing LSMO:NiO composite VAN thin films, out of plane lattice parameters $a_\perp$ (error ± 0.008 Å) of LSMO and NiO with % out of plane strain $\varepsilon_\perp$ (error ± 0.02%), temperatures ($T_2/T_1$), 15 kOe field cooled in-plane exchange bias field ($H_{EB}$), coercivity ($H_C$) and $T_{MI}$.

| Sample | NiO $a_\perp$(Å) | Strain NiO $\varepsilon_\perp$ (%) | LSMO $a_\perp$(Å) | Strain LSMO $\varepsilon_\perp$ (%) | LSMO $T_2/T_1$ (K) | $H_{EB}/H_C$ (Oe) | $T_{MI}$ (K) |
|---|---|---|---|---|---|---|---|
| ST7 | 4.092 | −2.10 | 3.887 | +0.46 | 196/242 | 43/363 | 240 |
| LA7 | 4.086 | −2.25 | 3.880 | +0.41 | 202/243 | 47/400 | 225 |
| ST14 | 4.060 | −2.86 | 3.902 | +0.85 | 133/230 | 210/555 | — |
| LA14 | 4.067 | −2.70 | 3.898 | +0.72 | 135/226 | 186/460 | — |
| ST25 | 4.09 | −2.15 | 3.890 | +0.54 | 180/265 | 55/195 | 210 |
| LA25 | 4.08 | −2.39 | 3.892 | +0.59 | 165/263 | 80/278 | 235 |
| Bulk | 4.18 | —— | 3.870 | —— | –/350 | —— | — |

To study the exchange bias effect in the composite thin films, the samples are cooled from 360 K (in the paramagnetic state of samples) to 5 K under a 15 kOe applied magnetic field. In the M-H hysteresis recorded at 5 K, we observe an exchange bias effect for all samples. To exclude the possibility of any artefacts we have also recorded the hysteresis at 5 K after field- cooling in a reversed field of -15 kOe FC. For LA7, LA14 and LA25 samples the ±15 kOe FC M-H at 5 K are shown in Figures. 4(b), (c) and (d) respectively. In the granular composite thin film LA7, the contact area between the two phases is smaller with smaller interfacial coupling. Therefore, as expected, the exchange bias in LA7 sample is very small (50 Oe), whereas the suppression of saturation magnetization is higher in samples for s =14 shots (LA14) because of the higher interfacial disorder arising from atomic disordering and magnetic frustration due to competing magnetic interaction at the interface.[33,39,44] For finer microstructures, the possibility of the pinning of uncompensated magnetic moments at the disordered interfaces is more favourable due to larger interfacial contact area (higher number of pinning sites).Therefore, we observe the highest coercivity $H_C$ and $H_{EB}$ values for nano-columnar (ST14) and dense nano-maze (LA14) microstructures.

Figure. 4(e) represents the thermal evolution of the exchange bias ($H_{EB}$) for the LA14 sample. The hysteresis loops at different temperature values (5 to 100 K) were recorded after field cooling under 15 kOe FC from 360 K. As shown in Figure 4(f)/4(g) for LA14/LA25, with increasing temperature the EB field rapidly decays in an exponential fashion and finally reduces to zero at ∼ 75 K / 78 K which is known to be the conventional blocking temperature ($T_B$). The coercivity also exhibits a continuous drop with increasing temperature as previously reported for the $La_{0.75}Sr_{0.25}MnO_3$/LaNiO$_3$ thin film system.[45] The calculated values of exchange bias [$H_{EB}= |(H_A + H_B)|/2$] and coercive field [$H_C= |(H_A - H_B)|/2$] for all samples are tabulated in Table 1 where $H_A$ and $H_B$ are the reverse and forward coercivity, respectively.

Most interestingly, the exchange bias is also observed in the out-of-plane (OP) direction as shown in Figure 5(a). The orientation of the sample surface and the magnetic field direction (⊥ to the sample surface) are shown in the inset of Figure 5(a). In-plane and out- of-plane 1.5 Tesla field cooled magnetic hysteresis of LA14 at 5K are shown in the Figure S2 in the Supporting Information. A significant magnetic anisotropy in the VAN sample results in higher exchange bias (230 Oe) and larger coercivity (830 Oe) in the OP direction compared to the corresponding values in the in- plane direction for the same LA14 sample, indicative of a strong perpendicular exchange bias (PEB).[34,37,46] These increased values of the EB field and Hc in the OP direction demonstrate that the EB effect dominates along the OP direction and the sample exhibits an anisotropic nature of the FM-AFM coupling at the interface.

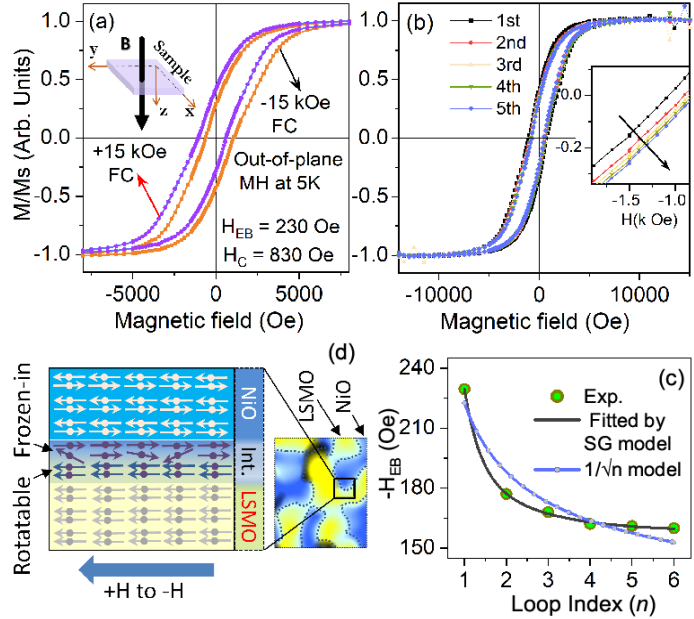

Figure. 5: (Color online) (a) OP hysteresis loop of sample LA14, recorded at 5 K after ±15 kOe field cooled cooling from 360 K. Inset represents the direction of applied magnetic field (B) and sample surface during the OP measurements. (b) Training effect at 5 K after 15 kOe field cooled cooling from 360 K. (c) Calculated $H_{EB}$ as the function of loop index 'n' and best fitting with $1/\sqrt{n}$ model and spin glass (SG) model. (d) Left bottom panel show the schematic representation of frozen-in and rotatable spins at the disorder interface during the training.

In order to understand the exchange bias and its origin, we have carried out measurements of the OP training effect at 5 K. The training effect refers to the phenomenon by which the EB is reduced during consecutive loop cycling $n$ due to interfacial spin rearrangement and relaxation. Six consecutive hysteresis loops after cooling in a field of 15 kOe to 5 K were measured, as shown in Figure 5(b). The calculated exchange bias field as a function of loop index $n$ is shown in Figure 5(c). First, the data is fitted by $1/\sqrt{n}$ model, typically used to describe the training of a magnetically ordered FM/AFM interface.[47] As can be seen in Figure 5(c), this fit does not describe our experimental data well. This is not surprising as the model mainly explains the training of a magnetically ordered FM/AFM interface:[48]

Therefore, we have chosen to attempt fitting of our data to the spin glass (SG) model of exchange bias.[49] According to this model, two types of antiferromagnetically (AFM) coupled NiO uncompensated spins (frozen and rotatable) exchange-couple with the ferromagnetic (FM) LSMO at the disordered interfaces, leading to significantly different relaxation rates: during the training, rotatable spins rotate with faster relaxation rate as compared to frozen spins. The data is fitted by equation:

$$H_{EB}^n = H_{EB}^\infty + A_{fz}exp^{(-n/P_{fz})} + A_{rt}exp^{(-n/P_{rt})}$$

Where, $H_{EB}^\infty$ is the EB field after sweeping the magnetic field for infinite times, $A_{fz}$, $P_{fz}$ and $A_{rt}$, $P_{rt}$ are the parameters related to the relaxation process in frozen and rotatable spins respectively at disordered FM/AFM interface. The fitting parameters ($H_{EB}^\infty$ = 158 ± 10 Oe, $A_{fz}$ =590 ± 22 Oe, $P_{fz}$ = 0.38 ±0.01 and $A_{rt}$=54 ± 5 Oe and $P_{rt}$ = 1.6 ±0.2) shows that when magnetic field switch from +H to –H in training the rotatable spins relax four times ($P_{rt}/P_{fz}$ ~ 4.2) faster as compared to frozen spin at the disordered interface as shown in the schematic in the left bottom penal of Figure 5(d).

**Transport properties**

In manganite based VAN thin films, strain engineering and microstructure is seen to play a very crucial role for tuning the magneto electrical transport properties of samples.24 Therefore, magneto transport properties as a function of temperature of these VAN thin films of different microstructures are investigated by resistivity measurements. Figure 6 shows the in-plane temperature-dependent behaviour of resistivity ρ(T) of LA7, LA14, and LA25 composite thin film samples in zero magnetic field and under the application of 80 kOe. The LA25 (ε = 0.59 %) composite thin film with nano-maze microstructure, shows a clear insulator-to-metal transition TMI at 230 K under 0 Oe magnetic field [Figure 6(c)]. At low temperatures below 70 K, the sample shows semiconducting behaviour with an upturn in resistivity. In contrast, the granular sample LA7 (ε = 0.41 %) shows a $T_{MI}$ at 225 K and a resistivity which is increased by one order of magnitude due to their granular microstructure with large grain boundaries [Figure 6(a)].

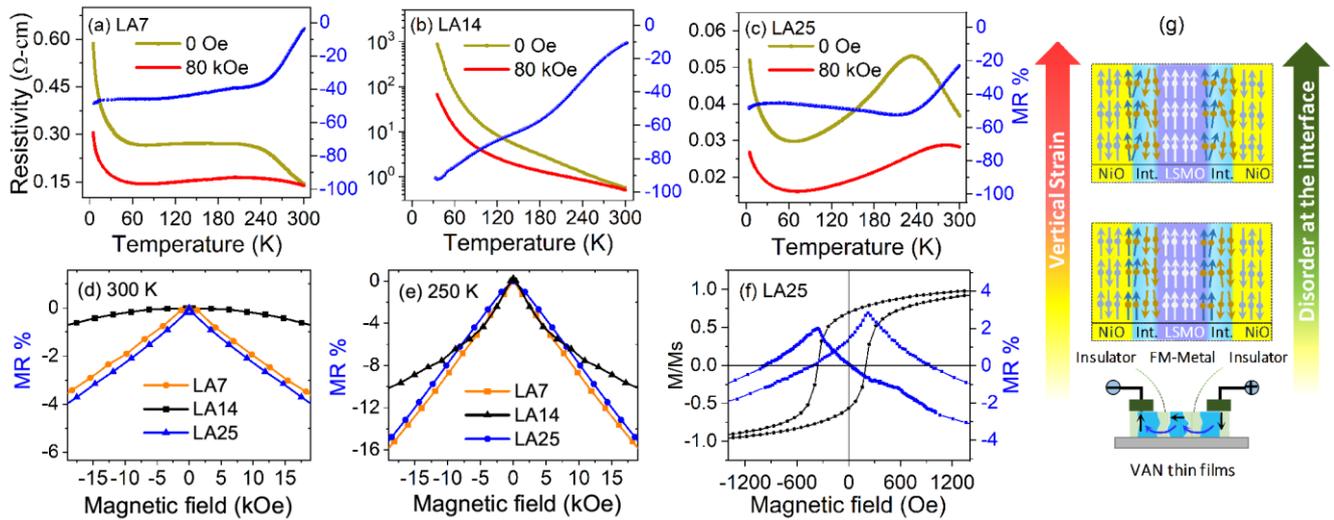

Figure 6: (a), (b) and (c) show the temperature-dependence of the resistivity in 0 and 80 kOe in-plane applied magnetic field, for samples LA7, LA14 and LA25 respectively. (d) and (e)show the LFMR of LA7, LA14 and LA25 composite thin films at 300 K and 250 K respectively. (f) +15 kOe field cooled MR and M-H of LA25 at 5 K. (g) Schematic of VAN heterostructure and LSMO(metallic):NiO (insulating) interface with disordered spins. Disorder increases with increasing vertical strain in the VAN thin film. Blue and black arrows in the VAN thin film figure (bottom) represent the path of conduction electrons across the vertical metal-insulator interface boundary.

Furthermore, it is interesting to note that in the LA14 sample, LSMO bears the highest strain (OP tensile strain) imposed by nearby NiO (OP compressive strain). The dense microstructure and high density of interface phase boundaries creates a series of LSMO/NiO/LSMO or FM-metal/insulator/FM- metal channels [as shown in the schematic Figure 6(g)], which leads to an enhancement in the electrical resistivity. Therefore, composite thin films LA14 (dense nano-maze, ε = 0.72%) and ST14 (nano columnar, ε = 0.85%) lose their metallic conductivities and show a huge (few orders of magnitude higher) enhancement in the resistivity and semiconducting behaviour without any metal-insulator transition [Figure 6(b)]. The enhancement in the resistivity and the low temperature upturn can be attributed to the presence of the highly resistive NiO phase boundaries with a large number of grains and increased disorder.[35,36] This highly resistive NiO phase obstructs the spin alignment and increases the tunnelling barrier height be- tween the neighbouring magnetic grains.[50] It should be noted that when the magnetic field 80 kOe is applied, the M-I transition of LSMO phase increases to 280 K in LA25 sample, whereas in LA7 sample, the M-I transition enhances to 265 K. Interestingly, even in the LA14 sample, which displays semiconducting behavior at all temperatures, a huge change in resistivity under the application of a magnetic field is observed, which is suggestive of a tunneling type of conduction between LSMO grains via insulating NiO grains.

The temperature dependent magnetoresistance MR % = [$R_T$ (H)-$R_T$ (0)/$R_T$ (0)] ×100 is shown in Figures 6 (a-c) under an applied field of 0 and 80 kOe. The MR % values of the LA25 thin film increase gradually with decreasing temperature and reach a maximum value of 52 % at 215 K under the applied field of 80 kOe. Interestingly, below 215 K, the MR % appears to be rather constant and steadily settles to 45 % at 10K. Similarly, the LA7 sample reveals a modest variation in MR % from 36 % at 240 K to 48 % at 10 K. Such variation in MR % to rather constant values over a wide temperature is known to be highly desirable for the application in spintronic devices. The LA14 (dense nano-maze) sample manifests semiconducting behaviour with decreasing temperature and shows the highest MR % values up to 92 % at 35 K.

More importantly, these composite VAN thin films exhibit a larger low-field magnetoresistance (LFMR) response near the room temperature as compared to other perovskite based nano composite thin film such as (LSMO)0.5:($CeO_2$)0.5.[51] Magnetoresistance as a function of magnetic field at 300 K and 250 K are shown in the Figure 6(d) and (e), respectively. In Figure 6(d), it is shown that nano-maze and nano-columnar samples exhibits more than 2 % LFMR (H ≤ 1T) at room temperature, while more than 7 % LFMR is observed for all three nanocomposite thin films at 250 K. For the LA25 sample, Figure 6(f) dis- plays the 15 kOe field-cooled (sample cooled from 330 K) MR % vs applied magnetic field and M-H hysteresis recorded at 5 K. The coercive fields obtained from the M-H curve and magnetoresistance MR % coincide[52] and both represent an almost similar shift in the field value, which again validates the presence of exchange bias effect phenomena in VAN thin films. These observations also clearly reveal that the introduction of a microstructure offers a new degree of freedom to tune the interface phase boundary density, disorder and the strain states which all play a crucial role for the control and the definition of the LFMR and magneto transport-behaviour of the composite thin films.

**Electronic structure**

To investigate the microscopic origins of the magnetic and transport properties, we further focussed on the study of the charge degrees of freedom. The information on the Ni/Mn valence states along with the hybridization between Mn/Ni-3d and O-2p orbitals in VAN thin films were deduced by the elemental site-selective near edge X-ray absorption spectroscopy (XAS) measurements at Mn and Ni $L$-, La $M$- and O $K$-edges, respectively as shown in Figure 7.

XAS spectra recorded at Mn $L$-edge for LA7, LA14 and LA25 samples along with LSMO thin film reference are shown in Figure 7(a). The Mn $L$-edge X-ray absorption spectra reveal two broad spin orbit split broad multiplets: $L_3$ at 642 eV photon energy resulting from a transition of Mn-$2p_{3/2}$ →3d and $L_2$ at 653 eV photon energy resulting from the Mn-$2p_{1/2}$ → 3d dipole transition. In the Mn $L_3$-edge of mixed-valence manganite, a shoulder-like feature corresponding to $Mn^{3+}$ appears at 640.6 eV, 1.5-1.6 eV lower than $Mn^{4+}$ main peak (642.15 eV).[53] When the strain value increases from 0.41 % to 0.85 % in the LA7 to LA14 samples, this distinct $Mn^{4+}$ feature in the $L_3$-edge (keeping the $L_2$ -edge position at fixed energy for better guide to eye) is gradually shifted towards higher photon energy by about 0.8 eV. This suggests that the $Mn^{4+}$ ion content increases with OP tensile strain.

In addition to this mentioned chemical shift, the increase of $Mn^{4+}$ content with increased OP tensile strain is evidenced by the systematic decrease in the spectral weight of the $Mn^{3+}$ high spin state feature[54] at photon energy 640.4 eV [shown by the circle in the inset if Figure 7(a)] along with the reduction of $L_3/L_2$ ratio ($L_3/L_2$ = 1.41, 1.38, 1.33, 1.34 for LSMO reference, LA7, LA14, LA25 samples, respectively). These observations reveal a reduction in the $Mn^{3+}$ fraction to fulfil the charge neutrality requirement. Recently Chandrasena et al.[9] reported an increase in $Mn^{3+}$ concentration associated with oxygen vacancy controlled by coherent in-plane tensile strain in $CaMnO_3$ thin films. In the present study, the composite thin films are prepared at the same OPP (250 mTorr) and on the same substrate (LA7, LA14, LA25 sample) thus, oxygen vacancy induced effects are minimal and similar for all the samples. Therefore, the observed effects are mainly attributed to the microstructure-induced strain, which leads to the modified $Mn^{4+}$ concentration and interfacial hybridization between Mn/Ni $3d$ and O $2p$ orbitals.

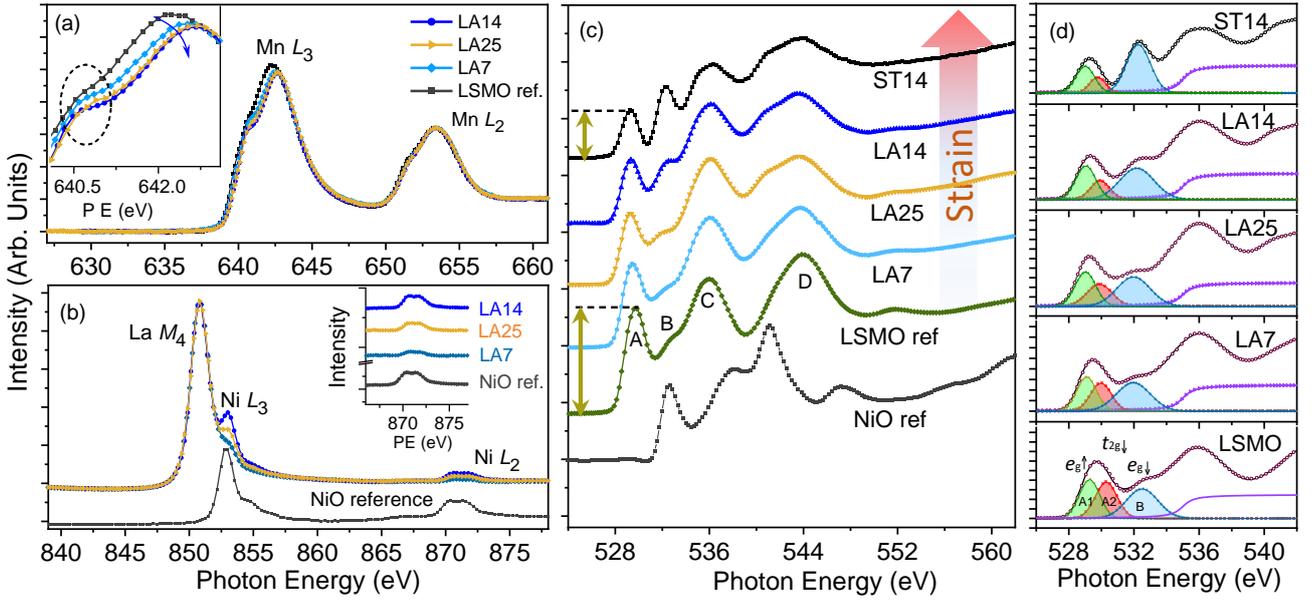

Figure 7: Room temperature pre-post edge corrected (a) XANES spectra of LA7, LA14, and LA25 along with LSMO reference recorded at Mn -edge, inset shows the close view of Mn -edge. (b) XANES spectra of LA7, LA14, and LA25 along with NiO (Ni 2+) reference at La $M_4$ + Ni $L$-edge, inset show the close view of Ni -edge. (c) O $K$ -edge of LA7, LA14, and LA25 along with LSMO and NiO reference samples. (d) room temperature O $K$ -edge spectra (o symbol) fitted by the combination of Gaussian and arctangent function for composite thin films and LSMO reference

Similar to the Mn $L$-edge X-ray absorption spectra, the Ni $L$-edge spectra also display spin orbit split multiplets $L_3$ and $L_2$ at 853 eV and 871 eV photon energy, respectively. However, in contrast to the Mn $L$-edge spectra, these spectra do not allow us to deduce the Ni valency from the Ni $L_3$-edge due to the overlap of the signal with that of the $M_4$-edge at 850.8 eV. Unfortunately, the intensity of the Ni $L_2$-edge is weak, but the Ni valency can nevertheless be estimated from the unique lineshapes of $Ni^{2+}$ and $Ni^{3+}$ at the Ni $L_2$-edge. The comparison of room temperature Ni -edge spectra [Inset of Figure 7(a)] accompanied by temperature dependent spectra of the Ni $L_2$-edge (from 320 K to 95 K) of the maximally strained LA14 sample with the simulated spectra of $Ni^{3+}$ and $Ni^{2+}$ shown in Figure 7(b) allow us to conclude that the Ni ion exists exclusively in its 2+ valence state in the composite thin films. These observations with temperature down to 95 K exclude the possibility of any charge transfer between Ni and Mn and clearly manifest that the transition appearing at temperature $T_2$ in the magnetization behavior is not related to a charge transfer effect.

Oxygen *K*-edge spectra uniquely reveals the variation in hybridization strength between oxygen and transition metal states (Mn/Ni 3*d* orbitals) via transition from the O-1*s* to O-2*p* states under the dipole allowed selection rule ($\Delta L = \pm 1$) and quantify the density of unoccupied states in O-2*p* levels.[55] Figure 7(c) shows the normalized O *K*-edge XANES spectra of LA7, LA14, LA25 and ST14 along with those of LSMO, NiO reference (100 nm thin films grown on STO under similar condition). Here, we find that the features A and B (shown by an arrow) from 528.5- 533.5 eV are attributed to unoccupied O-2*p* states[56] which are covalently intermixed with Mn-3*d* orbitals neighbouring a Ni-3*d* state[57]. Feature B is highly sensitive to the valence of Mn and strain. The broad bands represented by C and D are attributed to O-2*p* derived states hybridized with La-5*d* / Sr-4*f* and Mn- 4*sp* respectively.[53,58,59]

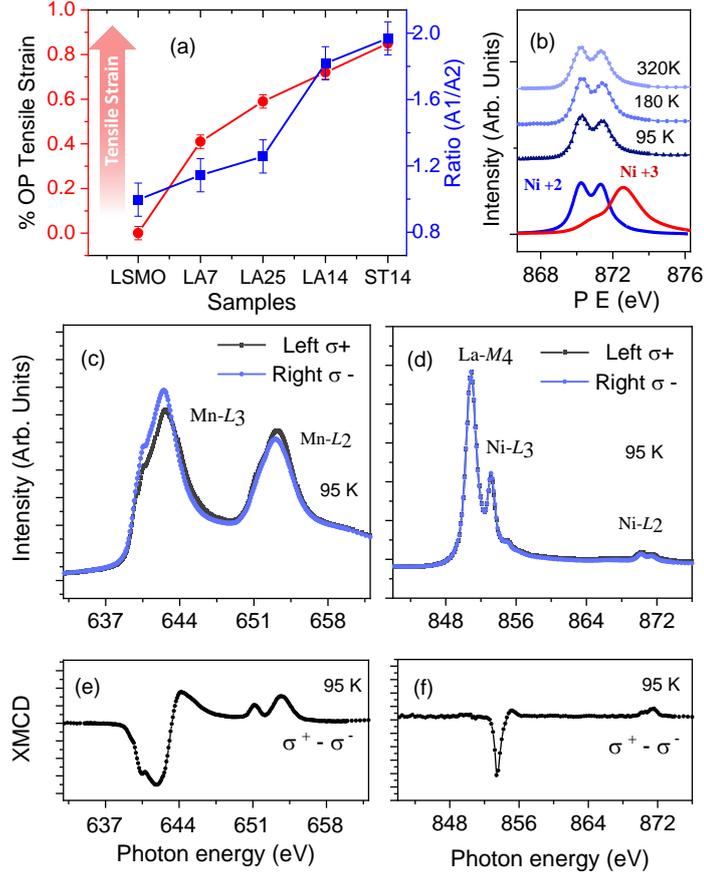

FIG. 8: (a) Variation of spectral weight ratio A1($e_g\uparrow$)/A2 ($t_{2g}\downarrow$) as a function of strain for the composite sample and LSMO reference. (b) Temperature dependent Ni -edge of LA14 sample along with simulated Ni edge for 2+ and 3+ valence state. Low temperature (95 K) left and right circularly polarized light XAS spectra of LA14 at (c) Mn *L*-edge in (d) and La $M_4$+ Ni -edge along with respective XMCD ( $\sigma^+ - \sigma^-$ ) at (e) Mn *L*-edge in and (f) La $M_4$+ Ni $L_{2,3}$-edge.

Here, we have focused on the feature A (A$_1$ as $e_g\uparrow$ and A$_2$ as $t_{2g}\downarrow$ as) and feature B ($e_g\downarrow$) because these features carry the main information about the hybridization between Mn/Ni 3*d* with O 2*p* states. We have fitted the spectra with the combination of an Arctan function as background and a series of Gaussian peaks[59] by using the software Athena. These $e_g$ and $t_{2g}$ features are distinctly visible after fitting the feature A (A$_1$ as $e_g\uparrow$ and A$_2$ as $t_{2g}\downarrow$ as) and feature B ($e_g\downarrow$). In the oxygen K-edge, the ratio of the spectral weight A1 ($e_g\uparrow$) and A2 ($t_{2g}\downarrow$) is close to 1 in the reference LSMO sample [Figure 7(d)]. As OP tensile strain is increased, a systematic growth in the spectral weight ratio A1/A2 is observed, which is plotted as a function of strain for composite samples as shown in Figure 8(a). This evolution of the spectral weight of A1 with strain, points towards an in- crease in the unoccupancy of the $e_g\uparrow$ state corroborated by the observation of an increase in the Mn$^{4+}$ ion concentration. These observations are fully consistent with the Mn $L$-edge results discussed above. The feature B at 533 eV is distinctly visible and its intensity monotonically increases with OP tensile strain. It is largest for the maximally strained samples (columnar ST14 and nano-maze LA14). The spectral weight of feature B is related to the hybridization of oxygen $2p$ with transition metal $3d$ orbitals and charge transfer effects from Mn to Ni as previously reported for the $Ni^{3+}$ systems LaMnO$_3$(n)-LaNiO$_3$(n)[60] and La$_{0.7}$Sr$_{0.3}$MnO$_3$/LaNiO$_3$[28] superlattice heterostructures. However, in the present samples, we exclude the possibility of charge-transfer from Mn to Ni or vice versa as the valence state of Ni is found to be +2 for all the samples as discussed above. Therefore, a systematic rise in spectral weight of feature B in the ST14 and LA14 samples directly indicates an enhanced degree of hybridization of oxygen $2p$ orbitals with mixed electrons of Mn or Ni $3d$ orbitals and a stronger coupling between Mn and Ni at the LSMO and NiO interface.

To get further insights into the relevant magnetic interactions between the LSMO and NiO phase at the interface, we performed the XMCD studies as shown in Figure 8 for the maximally strained sample LA14. At low temperature (95 K), NiO shows some ferromagnetic polarization due to a strongly coupled LSMO-NiO interface and which suggests the presence of a proximity effect driven ferromagnetism at the interface. The similar sign of XMCD at 95 K at Ni and Mn edge [Figures 8(e) and (f)], reveal that Mn$^{3+/4+}$ and Ni$^{2+}$ ions are ferromagnetically coupled at the interface either via modified superexchange or double exchange which could be the possible origin of the anomaly at $T_2$ in the M-T behaviour. In fact, a huge reduction in metallicity is observed for these highly strained LA14/ST14 and magnetically disordered samples,[36] such that FM superexchange[61] is favored compared to a double exchange mechanism.[62] The fine correspondence of our magnetization and XAS results divulges that the modification of the magnetic and transport properties is correlated to the strain induced strong interfacial hybridization between LSMO and NiO rather than charge transfer between Mn and Ni.

## Conclusions:

In summary, we have carried out PLD growth of LSMO:NiO self-assembled vertically aligned nanocomposite thin films of different microstructure on LaAlO$_3$(001) and SrTiO$_3$(001) substrates. The microstructures could be successfully modulated to re- veal granular, nano-maze and nano-pillar nano structures by controlling the laser shots without changing the LSMO-NiO phase fraction. DC magnetization reveals that magnetic properties such as magnetization and exchange bias in-plane and out-of-plane in the composite films are tuned by microstructure-induced disorder and out-of-plane strain. Training effect and XMCD measurements demonstrate that the observed exchange bias is disorder-induced and arises due to a pinning of NiO uncompensated moments at the disordered interface which is ferromagnetically coupled with LSMO. The nanostructure-tuned, microstructure-induced phase boundary significantly enhances the magnetoresistance (> 45% under 0 and 80 kOe) in a very broad temperature range (10-240 K) for the LA7 and LA25 samples and approaches > 90 % at low temperatures for LA14 sample. Furthermore, we have investigated the electronic properties of VAN LSMO:NiO composite thin films by performing XAS measurements, in which we observe systematic changes in the electronic structure across the vertical interface due to out-of-plane tensile strain induced Mn$^{3+}$/Mn$^{4+}$ content rather than due to a charge transfer effect. We demonstrate that the vertical strain in the heterointerfaces plays a crucial role for the control and the tuning of the magnetic and trans- port properties. Our results provide a new pathway for controlling the exchange bias and magneto-transport properties via nanostructure-driven, microstructure-induced vertical strain in oxides heterostructures, useful for the design of future advanced multifunctional devices, such as advanced magnetoresistive recording read heads, magnetic random access memory devices or other spintronic applications.

**Acknowledgments:**

The authors are thankful to R. Rawat for providing resistivity measurements. The authors are thankful to Dr. V. R. Reddy for providing RSM measurements. Gyanendra Panchal is thankful to Klaus Habicht for careful reading of and providing valuable comments on the manuscript. Gyanendra Panchal is thankful to Anupam Jana for help during EDX measurements. Authors also acknowledge Rakesh K. Sah for help during XAS measurements.

**Conflict of Interest:** The authors declare no conflict of interest.

# Supporting Information

# Synthesis and Characterization of Vertically Aligned La$_{0.7}$Sr$_{0.3}$MnO$_3$:NiO Nanocomposite Thin Films for Spintronic Applications

Gyanendra Panchal[1,2*], Anjali Panchwanee[2], Manish Kumar[3], Katharina Fritsch[1], Ram Janay Choudhary[2*], and Deodutta Moreshwar Phase[2]

[1] *Department of Methods for Characterization of Transport Phenomena in Energy Materials, Helmholtz-Zentrum Berlin für Materialien und Energie, D-14109 Berlin, Germany*
[2] *UGC DAE Consortium for Scientific Research, University Campus, Khandwa Road, Indore 452001, India*
[3] *Pohang Accelerator Laboratory, POSTECH, Pohang 37673, South Korea*

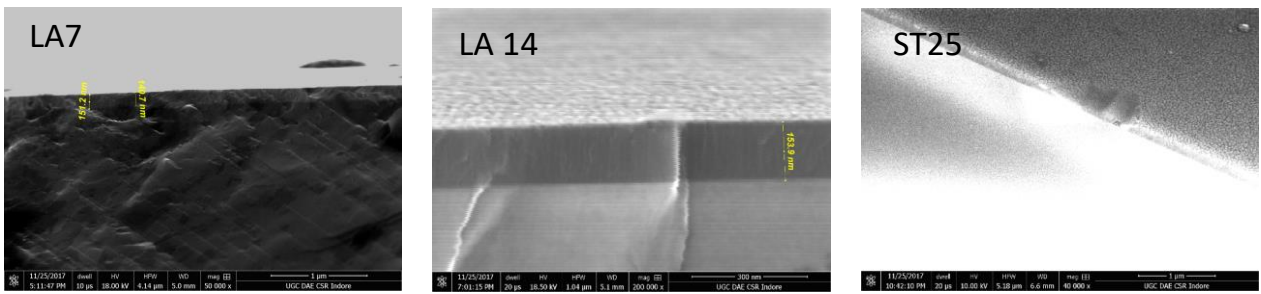

Figure S1. Cross sectional views of LA7, LA14 and ST25 (tilted at 60°) nanocomposite thin films.

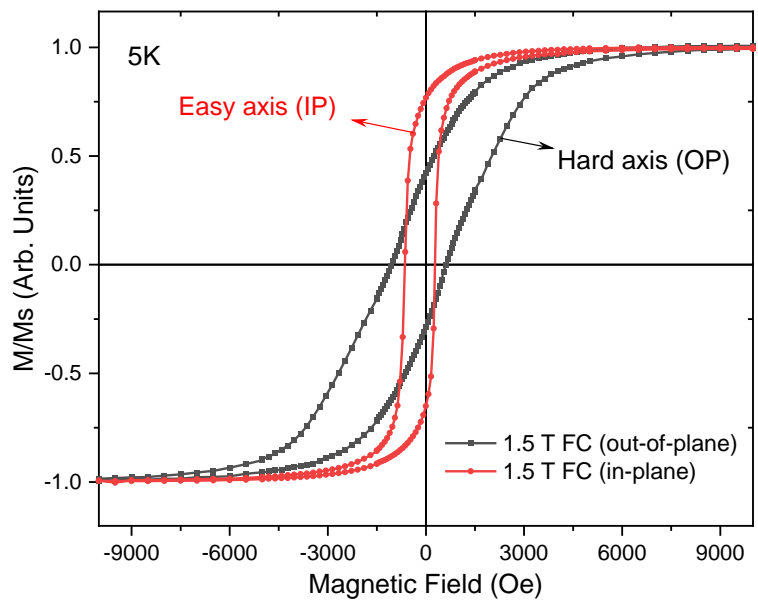

Figure S2. In-plane and out-of-plane 1.5 Tesla field cooled magnetic hysteresis behavior of LA14 composite thin film at 5K.